# Theoretical Prediction of Superconducting 2D Hydrogenated Janus $V_2CSH$ and $Nb_2CSH$ MXenes

Jubair Hossan Abir, M.R.U.S. Shapon, Md. Sharear Aman, Md. Kamrul Hassan, A.K.M. Azharul Islam, Saleh Hasan Naqib*

[1]Department of Physics, University of Rajshahi, Rajshahi 6205
*Corresponding author; Email: jsr_eda1.ru.ac.bd (Saleh Hasan Naqib)

## Abstract

Two-dimensional hydrogenated Janus MXenes represent a new frontier in superconducting materials, where asymmetric surface terminations enable tunable electronic and vibrational properties. In this work, we present the first comparative study of hydrogen/sulfur-terminated $V_2CSH$ and $Nb_2CSH$ monolayers using density functional theory and density functional perturbation theory. Both compounds are structurally and thermally stable, exhibiting isotropic in-plane elasticity and nonmagnetic multiband metallic states dominated by transition-metal *d* orbitals. The asymmetric S/H terminations generate nonequivalent surface potentials and intrinsic out-of-plane dipoles. Electron-phonon coupling constants of 0.39 ($V_2CSH$) and 0.53 ($Nb_2CSH$) yield superconducting transition temperatures of 2.36 K and 5.27 K, respectively, within the Allen-Dynes framework. Low-energy lattice vibrations (< 60 meV), primarily involving metal and sulfur atoms, account for ~77% of the coupling in both MXenes. These findings establish metal substitution as a design principle for tuning superconductivity in Janus MXenes and highlight $Nb_2CSH$ as a promising candidate for low-temperature superconducting applications.

## 1 Introduction

MXenes entered the two-dimensional (2D) materials literature through a specific chemical operation rather than mechanical exfoliation: selective removal of Al from $Ti_3AlC_2$ produced $Ti_3C_2$ nanosheets in 2011 [1]. The same strategy was soon extended to additional transition-metal carbides and carbonitrides [2]. MXenes are commonly written as $M_{n+1}X_nT_x$, where M is an early transition metal, X is C or N, and $T_x$ denotes surface species introduced or retained during synthesis [3]. $Nb_2C$ and $V_2C$ are not merely theoretical members of this family; Naguib et al. prepared both by selectively etching Al from $Nb_2AlC$ and $V_2AlC$, respectively [4]. Their chemically accessible surfaces and interlayer galleries are central to MXene behavior. In $Ti_3C_2$, for example, cation intercalation produced high volumetric capacitance, and dense conductive films could be processed from the etched carbide [5,6].

The termination layer is an important part of the electronic structure of an MXene rather than a passive surface decoration. Khazaei et al. compared F-, O-, and OH-terminated $M_2C$ and $M_2N$ monolayers and obtained termination-dependent metallic, semiconducting, and magnetic states [7]. A later calculation from the same group identified exceptionally low work functions for several OH-terminated carbides and nitrides [8]. Hart et al. connected surface defunctionalization directly with increased electronic conductivity in $Ti_3C_2T_x$ and showed that intercalation can shift the temperature dependence of transport between metallic and semiconductor-like regimes [9]. A broader range of surface chemistries became experimentally accessible when Kamysbayev et al. used molten-salt reactions to install or remove ligands including S, Se, Cl, Br, and Te. The same study showed that ligands change lattice dimensions and the superconducting response of Nb-based carbide MXenes [10]. Covalent and non-covalent functionalization routes now provide additional ways to alter surface composition beyond the termination mixture inherited from etching [11].

Out-of-plane asymmetry can be introduced in more than one way and the distinction matters for H/S-terminated carbides. Frey et al. examined surface-engineered magnetic MXenes and predicted that the magnetic exchange and anisotropy can be modified by surface chemistry together with an applied electric field [12]. Akgenc introduced Janus character through inequivalent transition-metal sublayers in M′MC and $M'MCT_2$ models and those structures displayed termination-dependent metallic, half-metallic, and semiconducting states, with dynamical and thermal stability assessed for selected compositions [13]. Li et al. treated a different type of asymmetry in $Ti_3C_2F_xO_{2-x}$, where the two surfaces carry unequal F/O environments and resolved changes in both electronic structure and mechanical response as the termination composition and disorder were varied [14]. These calculations do not directly describe H/S-terminated $M_2C$, but they show why two chemically inequivalent faces cannot, in general, be represented by a symmetrically terminated reference sheet.

Sulfur termination has a more direct connection to $Nb_2C$ and $V_2C$. In a first-principles screening of 20 sulfur-terminated early-transition-metal carbides and nitrides, Yang et al. found $V_2CS_2$ and $Nb_2CS_2$ among the systems that remained dynamically stable in phonon calculations and finite-temperature molecular dynamics [15]. A related experimental S-Nb-C-Nb-S slab has also been reported: Sakamaki et al. prepared single-phase 1T-$Nb_2S_2C$ by extracting Cu from 1T-$Cu_{2/3}Nb_2S_2C$ while retaining the layered carbosulfide framework [16]. That compound is not a MAX-derived MXene, so it should not be treated as experimental confirmation of $Nb_2CS_2$, but it establishes that the same elemental stacking motif is chemically realizable. For $V_2C$, Yan et al. compared O- and S-terminated sheets with F- and OH-terminated counterparts and reported metallic $V_2CO_2$ and $V_2CS_2$ together with favorable Li adsorption and diffusion characteristics [17]. Champagne et al., working with pristine, O-terminated, and OH-terminated $V_2C$, separately mapped how termination changes the electronic bands, phonon spectra, and dynamical stability [18].

The experimental record for $V_2C$ also shows that the underlying carbide is sensitive to preparation conditions. Wu et al. found that, among the fluoride/HCl mixtures they tested in an open reaction environment, NaF + HCl produced high-purity $V_2C$ and they quantified its thermal stability in inert and oxidizing atmospheres [19]. A later comparison of open and closed etching showed that a closed hydrothermal environment broadens the usable chemistry to LiF + HCl and can improve the purity and layered morphology of the product [20]. For $Nb_2C$, complementary studies have focused on termination-specific electronic descriptors. Ghasemy et al. calculated F-, Cl-, S-, and mixed-terminated $Nb_2C$ surfaces and linked the termination identity to changes in work function, Nb *d*-band position, adsorption, and molecular dissociation barriers [21]. Liu et al. extended the termination set to O, S, Se, Te, halogens, and OH and in their calculations S and Se changed $Nb_2C$ from metallic to semiconducting while several terminations also shifted the work function substantially [22]. The response of $Nb_2C$ and $V_2C$ to a ligand therefore depends on the metal center as well as on the ligand itself.

Hydrogen functionalization provides another relevant precedent. Xue et al. designed hydrogenated $Ti_2N$ MXenes, $Ti_2NH_2$ and $Ti_2NH_4$, and found that hydrogenation changed the states near the Fermi level and the electron-phonon response with superconductivity predicted for the hydrogen-rich phase [23]. These hydrogenated phases contain no sulfur and are therefore chemically distinct from an H/S Janus carbide. Wan et al. experimentally produced metallic Janus MoSH by controlled $H_2$-plasma treatment of $MoS_2$ [24]. MoSH belongs to the transition-metal chalcogenide family rather than the MXene family, yet it provides direct experimental evidence that H can replace the chalcogen on only one face of a two-dimensional layer. Theoretical screening of Janus transition-metal sulfhydrates subsequently identified several metallic, phonon-stable candidates with electron-phonon-mediated superconductivity [25]. The closest carbide analogue is $Ti_2CSH$. Seeyangnok and Pinsook reported in 2026 that hydrogenation of $Ti_2CS_2$ to $Ti_2CSH$ yields a dynamically stable Janus MXene with electron-phonon-mediated superconductivity [26].

Despite these advances, one specific comparison remains unexplored. Sulfur-terminated $V_2C$ and $Nb_2C$ have been examined directly [15,17,21,22]. Hydrogenated MXenes and H/S Janus chalcogenides provide separate hydrogen-related precedents [23–25], and $Ti_2CSH$ provides an H/S-asymmetric carbide analogue [26]. None of those studies, however, addresses the same H on one-face/S on the other termination pattern on both $V_2C$ and $Nb_2C$. Because the two parent carbides respond differently to surface chemistry, the Ti-based result cannot be transferred to V- and Nb-centered sheets by simple chemical analogy. In this work, $Nb_2CSH$ and $V_2CSH$ are considered with H/S-asymmetrically terminated $M_2C$ monolayers. We have examined using first-principles calculations to assess their structural and thermal stability, electronic structure, lattice dynamics, and electron-phonon coupling relevant to superconductivity. This direct comparison links established $V_2C$/$Nb_2C$ termination chemistry with the emerging H/S Janus-carbide motif without assuming that the behavior of $Ti_2CSH$ carries over to V- and Nb-centered systems.

## 2 Computational Methodology

First-principles calculations were carried out within density functional theory (DFT) using the Vienna Ab-initio Simulation Package (VASP) [27,28]. The electron-ion interactions were treated using the projector augmented-wave (PAW) method [29], while the exchange-correlation energy was described by the Perdew-Burke-Ernzerhof (PBE) functional within the generalized gradient approximation (GGA) [30]. Under periodic boundary conditions, a unit cell of each $V_2CSH$ and $Nb_2CSH$ monolayer was separated from its periodic images by a vacuum region exceeding 25 Å along the out-of-plane direction. Following convergence tests, a plane-wave kinetic-energy cutoff of 500 eV and a Brillouin-zone sampling resolution of $2\pi \times 0.02$ Å$^{-1}$ were adopted. Because of the intrinsic out-of-plane polarization, a self-consistent dipole correction was applied with the dipole reference positioned in the vacuum region to compensate for the difference between the vacuum levels on the two sides of the monolayer. Long-range dispersion interactions were included through the DFT-D3 correction [31]. The total-energy and force convergence criteria were set to $10^{-6}$ eV and 0.02 eV/Å, respectively.

Ab initio molecular dynamics (AIMD) simulations were performed to evaluate the thermal stability of both monolayers. The simulations were conducted in the canonical ensemble using a Nosé-Hoover thermostat (NVT). For each system, a $2 \times 2 \times 1$ supercell was simulated at 300 K for 5.0 ps using a time step of 1.0 fs.

Lattice-dynamics and electron-phonon coupling (EPC) calculations were performed using Quantum ESPRESSO [32] within density-functional perturbation theory (DFPT) [33]. Optimized norm-conserving Vanderbilt pseudopotentials from the PseudoDojo library [34] were employed with a total-energy convergence threshold of $10^{-10}$ Ry and a plane-wave cutoff of 80 Ry. The phonon properties of $V_2CSH$ and $Nb_2CSH$ were calculated using $12 \times 12 \times 1$ $q$-point meshes. The Born-Huang invariance and Huang equilibrium conditions were imposed on the harmonic interatomic force constants to ensure physically consistent lattice dynamics [35].

# 3 Results and discussion

## 3.1 Crystal Structure and Asymmetric Charge Transfer

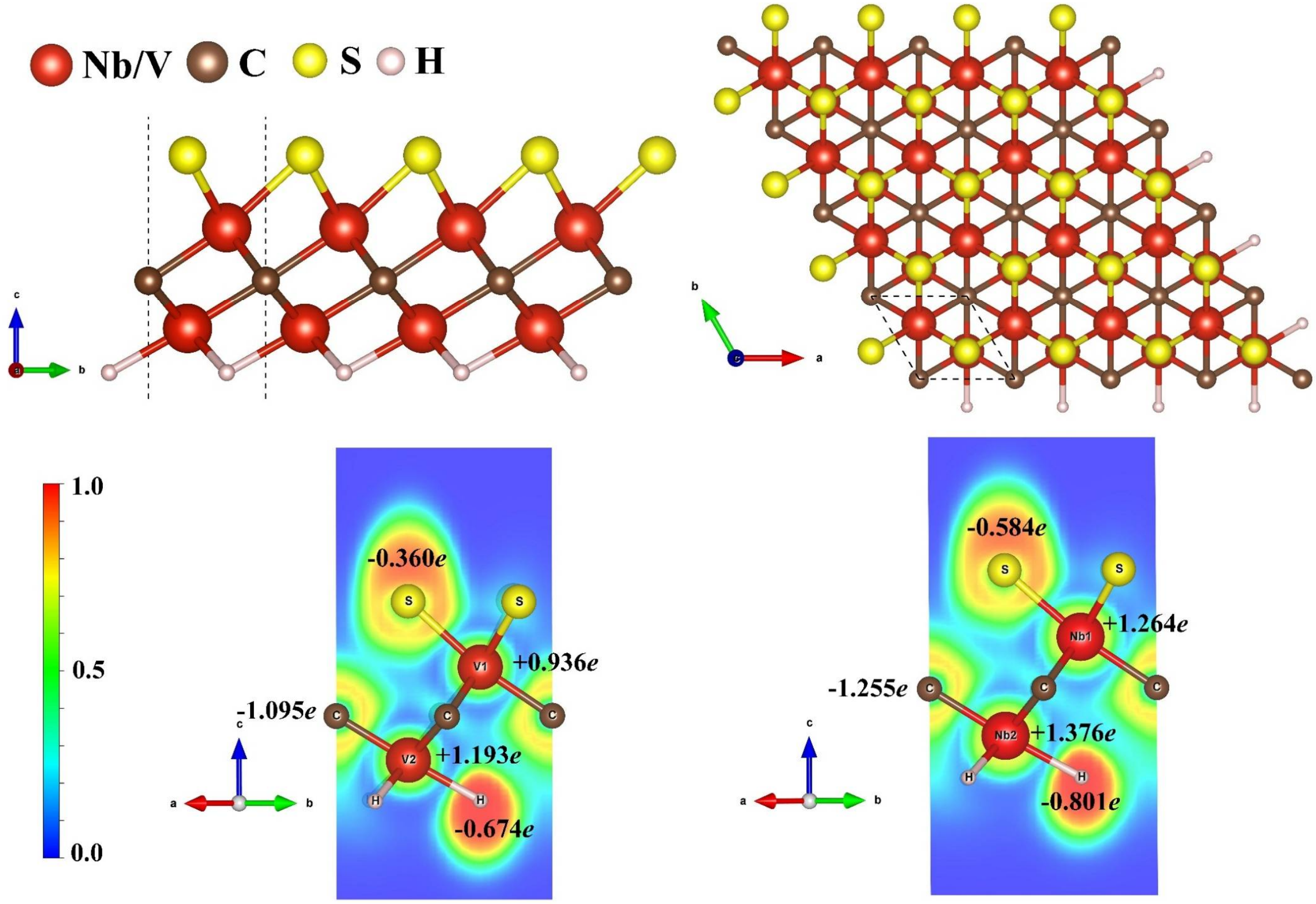

**Figure 1**: Structural configurations and electron localization function (ELF) maps of Janus $V_2CSH$ and $Nb_2CSH$ monolayers. Upper panels: Side (left) and top (right) views showing S and H terminations on opposite surfaces. Red, brown, yellow, and pink spheres represent V/Nb, C, S, and H atoms, respectively. Dashed lines show the unit cell. Lower panels: ELF maps of $V_2CSH$ (left) and $Nb_2CSH$ (right), ranging from 0 (blue) to 1 (red). Annotated Bader charges are expressed in units of e, with positive and negative values indicating electron loss and gain, respectively.

**Figure 1** presents the crystal structures, electron localization function (ELF) maps and Bader charges of Janus $V_2CSH$ and $Nb_2CSH$ monolayers. Each primitive cell contains five atoms: two V or Nb atoms, one C atom, one S atom, and one H atom. The carbon layer is sandwiched between two transition-metal sublayers, with S and H terminating opposite surfaces. The crystallographic models are consistent with the trigonal space group *P3m1* (No. 156). Both structures lack inversion symmetry and a horizontal mirror plane, while retaining a threefold rotation axis and vertical mirror planes. This architecture shares the asymmetric S/H termination motif investigated for Janus $Ti_2CSH$ [**26**].

The in-plane lattice constants are 2.968 Å for $V_2CSH$ and 3.191 Å for $Nb_2CSH$, corresponding to a 7.51% expansion upon replacing V with Nb. This stems mainly from the difference in the ionic size of the atoms involved. The monolayer thickness, defined as the perpendicular separation between the outermost S and H atomic planes, increases from 4.743 to 5.049 Å. The two transition-metal sublayers also exhibit different local bond lengths. The nearest V-C distances are 2.068 Å on the S-facing side and 2.009 Å on the H-facing side, compared with 2.223 and 2.171 Å for the corresponding Nb-C bonds. The V-S and V-H distances are 2.326 and 1.966 Å, respectively, whereas the Nb-S and Nb-H distances are 2.466 and 2.105 Å. Thus, Nb substitution expands both the in-plane lattice and the local coordination environment while retaining the Janus stacking arrangement.

The spin-polarized band structure and PDOS without SOC calculations were performed to verify possible magnetic ground state of both monolayers. The spin-polarized electronic structures show essentially coincident spin-up and spin-down bands along Γ-M-K-Γ, accompanied by spin-resolved densities of states for both monolayers (**Appendix A**, **Figure 9**). These results are consistent with a nonmagnetic character of both optimized structures.

The ELF maps in **Figure 1** provide a spatial description of electron localization. Both compounds exhibit pronounced localization around C, S, and H atoms. The distinct localization patterns at the S and H-terminated surfaces reflect their different electronic environments. Together with the Bader charges, this distribution supports a substantial charge-transfer component of the bonding. Bader charge analyses quantify the net atomic charge redistribution [**36**]. Using the convention that positive charges denote electron loss and negative charges denote electron gain relative to neutral atoms, the S-facing and H-facing V atoms carry charges of +0.936e and +1.193e, respectively. The corresponding C, S, and H charges are -1.095e, -0.360e, and -0.674e. In $Nb_2CSH$, the S-facing and H-facing Nb atoms carry +1.264e and +1.376e, while C, S, and H carry -1.255e, -0.584e, and -0.801e, respectively. Both transition-metal sublayers therefore donate electronic charge to the carbon framework and surface terminations. Carbon is the largest individual charge acceptor in each monolayer, and the negative H charge identifies an electron-enriched hydrogen termination.

The total positive Bader charge on the two transition-metal atoms increases from 2.129e in $V_2CSH$ to 2.640e in $Nb_2CSH$, representing approximately 24.0% greater charge transfer within this partitioning scheme. Therefore, $Nb_2CSH$ exhibits greater overall charge transfer but a smaller charge imbalance between its two transition-metal sites. These are distinct descriptors: the net out-of-plane dipole depends on the complete spatial distribution of ionic and electronic charge, including both terminations.

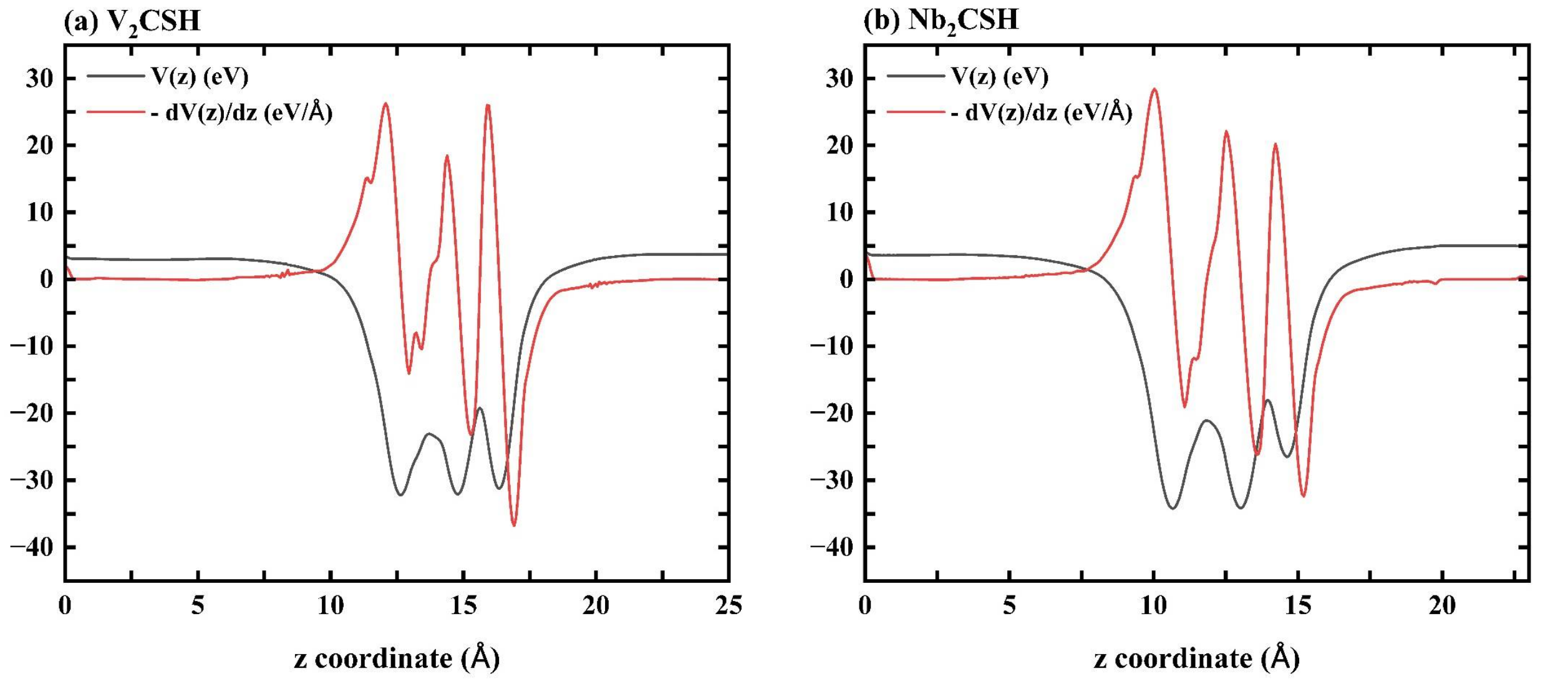

**Figure 2**: Planar-averaged electrostatic potential V($z$) (black) and corresponding electric field (-dV($z$)/d$z$) (red) along the out-of-plane direction of Janus (a) $V_2CSH$ and (b) $Nb_2CSH$ monolayers.

The vacuum plateaus on opposite sides of each monolayer occur at different potential levels, consistent with an intrinsic out-of-plane dipole associated with the chemically inequivalent S and H surfaces. Their separation is larger for $Nb_2CSH$, suggesting a more pronounced surface-potential asymmetry. A finite potential step across a dipolar slab is compatible with a vanishing electric field far from its surfaces [**37**]. Consequently, the nearly flat vacuum regions do not imply that the monolayer lacks an intrinsic dipole. Termination-dependent vacuum levels have likewise been reported for Janus $Y_2CClX$ MXenes (X = F, Br, I) [**38**].

## 3.2 Elastic Stability and Isotropic Mechanical Response

The linear in-plane elastic response of Janus $V_2CSH$ and $Nb_2CSH$ was evaluated using the stress-strain method. Both monolayers satisfy the elastic stability criteria, with $V_2CSH$ exhibiting modestly higher axial and shear stiffness. Their in-plane symmetry reduces the elastic response to two independent coefficients, $C_{11}$ and $C_{12}$. In Voigt notation, the stiffness matrix takes the form [**39**,**40**]:

$$C = \begin{pmatrix} C_{11} & C_{12} & 0 \\ C_{12} & C_{11} & 0 \\ 0 & 0 & C_{66} \end{pmatrix}. \quad (1)$$

where symmetry imposes,

$$C_{22} = C_{11}, \qquad C_{66} = \frac{C_{11}-C_{12}}{2}. \quad (2)$$

For an unstressed sheet, mechanical stability requires positive strain energy under any infinitesimal homogeneous in-plane deformation. For the elastic form above, the Born conditions reduce to [**40**],

$$C_{11} > 0, \; C_{11}^2 - C_{12}^2 > 0. \quad (3)$$

The calculated elastic constants satisfy both conditions for $V_2CSH$ and $Nb_2CSH$ (**Table 1**). The corresponding Voigt stiffness-matrix eigenvalues are (91.47, 182.94, 266.57) and (88.43, 176.85, 264.70) $Nm^{-1}$, respectively. Their positive signs provide an equivalent confirmation of elastic stability but dynamical and thermal stability require separate evidence.

**Table 1**: In-plane elastic constants and derived mechanical properties of $V_2CSH$ and $Nb_2CSH$. Elastic stiffnesses and moduli are in $Nm^{-1}$; ν and B/G are dimensionless.

| Monolayer | $C_{11}$ | $C_{12}$ | $C_{66}$ = G | E | B | ν | B/G |
|---|---|---|---|---|---|---|---|
| $V_2CSH$ | 224.76 | 41.81 | 91.47 | 216.98 | 133.28 | 0.186 | 1.457 |
| $Nb_2CSH$ | 220.78 | 43.92 | 88.43 | 212.04 | 132.35 | 0.199 | 1.497 |

The directional Young's modulus and Poisson's ratio were obtained from the in-plane compliance tensor [**39**]:

$$E(\theta) = \frac{C_{11}C_{22} - C_{12}^2}{C_{22}\cos^4\theta + A\cos^2\theta\sin^2\theta + C_{11}\sin^4\theta} \tag{4}$$

$$\nu(\theta) = \frac{C_{12}\cos^4\theta - B\cos^2\theta\sin^2\theta + C_{12}\sin^4\theta}{C_{22}\cos^4\theta + A\cos^2\theta\sin^2\theta + C_{11}\sin^4\theta} \tag{5}$$

Here, θ is the angle between the loading direction and the in-plane x axis and the auxiliary coefficients are:

$$\alpha = \frac{C_{11}C_{22} - C_{12}^2}{C_{66}} - 2C_{12}, \quad \beta = C_{11} + C_{22} - \frac{C_{11}C_{22} - C_{12}^2}{C_{66}}. \tag{6}$$

Applying **Equation(2)** gives $\alpha = 2C_{11}$ and $\beta = -2C_{12}$, removing the angular dependence. The resulting Young's modulus E, Poisson's ratio ν, two-dimensional bulk (areal) modulus B, and shear modulus G are:

$$E = \frac{C_{11}^2 - C_{12}^2}{C_{11}}, \qquad \nu = \frac{C_{12}}{C_{11}} \tag{7}$$

$$B = \frac{C_{11} + C_{12}}{2}, \qquad G = C_{66} = \frac{C_{11} - C_{12}}{2} \tag{8}$$

$V_2CSH$ has a Young's modulus of 216.98 $Nm^{-1}$ and a shear modulus of 91.47 $Nm^{-1}$, compared with 212.04 and 88.43 Nm-$^{1}$, respectively, for $Nb_2CSH$. Replacing Nb with V raises E by 2.3% and G by 3.4%, whereas B changes by only 0.7%. The larger relative change in shear stiffness indicates greater sensitivity of resistance to shape deformation to the metal species within the calculated linear response.

The calculated Young's moduli are below experimental benchmarks for graphene (340 $Nm^{-1}$) [**41**] and monolayer h-BN (289 ± 24 $Nm^{-1}$) [**42**]. Both lie within the interval corresponding to the reported monolayer $MoS_2$ modulus (180 ± 60 $Nm^{-1}$) [**43**]. These comparisons provide experimental context for the calculated in-plane stiffnesses.

The Poisson's ratios of $V_2CSH$ and $Nb_2CSH$ are 0.186 and 0.199, respectively. Both monolayers therefore contract transversely under uniaxial tension, with a smaller lateral response in $V_2CSH$. Their positive Poisson's ratios exclude in-plane auxetic behavior within this elastic regime. The corresponding B/G ratios, 1.457 and 1.497, describe the balance between areal and shear stiffness. Because the Pugh criterion was developed for bulk polycrystalline metals [44], these two-dimensional ratios alone do not establish ductility or brittleness with certainty.

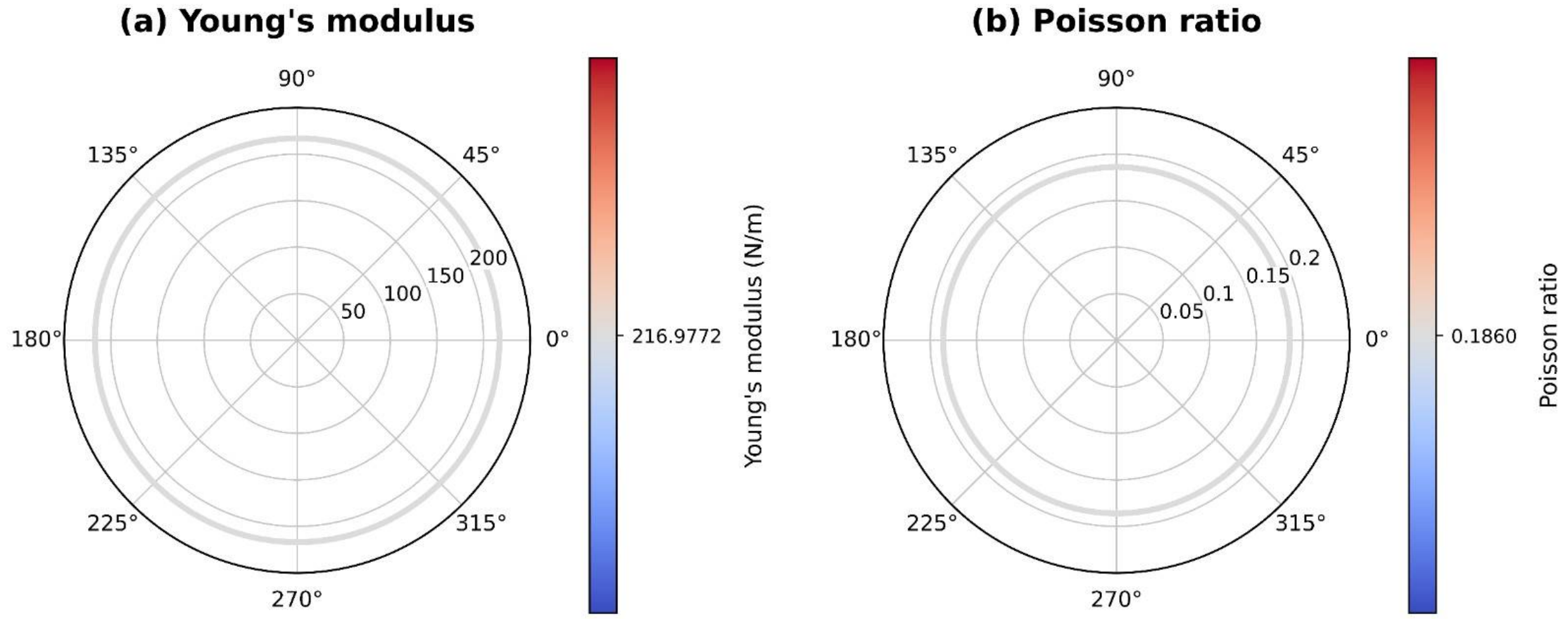

Figure 3: Angular dependence of (a) Young's modulus and (b) Poisson's ratio of $V_2CSH$. The circular profiles correspond to E = 216.98 N $m^{-1}$ and ν = 0.186.

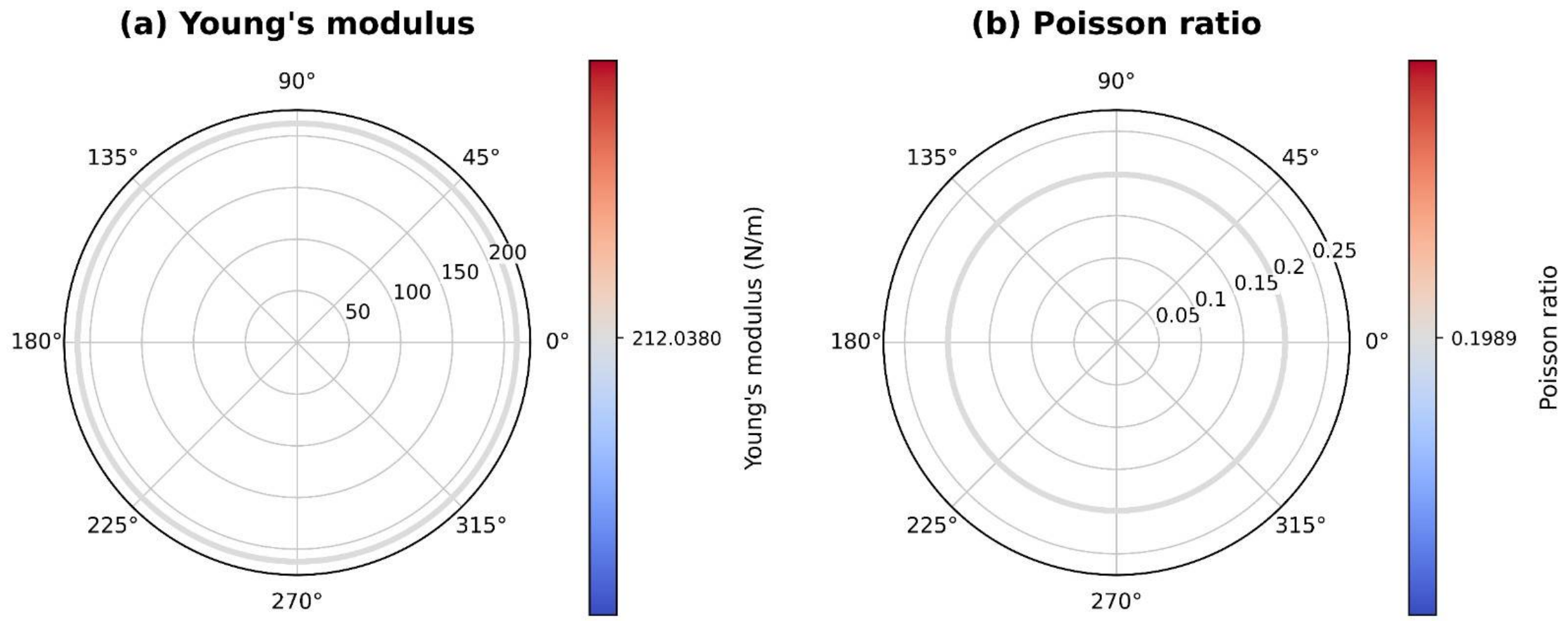

**Figure 4**: Angular dependence of (a) Young's modulus and (b) Poisson's ratio of $Nb_2CSH$. The circular profiles correspond to E = 212.04 N $m^{-1}$ and ν = 0.199.

The circular polar profiles of $V_2CSH$ (**Figure 3**) and $Nb_2CSH$ (**Figure 4**) follow directly from the angular invariance of E and ν. The directional extrema coincide, giving $E_{max}/E_{min} = G_{max}/G_{min} = 1$, consistent with two-dimensional elastic isotropy [39]. Thus, the chemically inequivalent Janus surfaces coexist with direction-independent in-plane linear elasticity. The composition changes the magnitude of the elastic response while preserving its isotropic character.

## 3.3 Energetic Favorability

The formation energies of the Janus $M_2CSH$ monolayers (M = V or Nb) were evaluated relative to selected elemental reference phases from the Materials Project [**45**]:

$$E_f = \frac{5E_{M_2CSH} - 2E_{M,bulk} - E_{C,bulk} - E_{S,bulk} - E_{H,bulk}}{5} \quad (9)$$

Where, $E_{M_2CSH}$ denotes the energy per atom of the relaxed Janus $M_2CSH$ monolayer (M = V or Nb), and $E_{M,bulk}$, $E_{C,bulk}$, $E_{S,bulk}$, and $E_{H,bulk}$ denote the energies per atom of cubic V or Nb ($Im\bar{3}m$), monoclinic C (C2/m), monoclinic S ($P2_1$), and orthorhombic H ($P2_12_12_1$), respectively.

The calculated formation energies are -0.41062 eV-atom$^{-1}$ for $V_2CSH$ and -0.51731 eV-atom$^{-1}$ for $Nb_2CSH$, equivalent to -39.62 and -49.91 kJ-mol$^{-1}$, respectively. These negative formation energies indicate energetically favorable formation relative to the selected elemental references. Further investigation is required to establish stability against various competing phases and assess the feasibility of experimental synthesis through methods such as chemical vapor deposition (CVD).

## 3.4 Thermal Stability

To assess the thermal robustness of the Janus $V_2CSH$ and $Nb_2CSH$ monolayers beyond harmonic phonon calculations and static elastic analyses, *ab initio* molecular dynamics (AIMD) simulations were performed at 300 K. By explicitly tracking thermally driven atomic motion, AIMD provides an atomistic assessment of structural integrity under finite-temperature perturbations, including possible lattice distortions and surface reconstructions. Each monolayer was modeled using a 2 × 2 × 1 supercell containing 20 atoms in the canonical (NVT) ensemble. The simulations were propagated for 5,000 steps with a time step of 1 fs, corresponding to a total duration of 5 ps.

As shown in **Figure 5(a, b)**, the total energies exhibit bounded fluctuations following the initial transient, while the smoothed temperature profiles remain close to the target temperature of 300 K. Comparison of the initial and final atomic configurations reveals preservation of the characteristic Janus framework, with no apparent structural collapse or surface reconstruction. The retention of this architecture under thermal excitation provides direct evidence of room-temperature structural stability over the simulated timescale. Together with the phonon and elastic results, these findings strongly support the structural robustness of both $V_2CSH$ and $Nb_2CSH$ under the investigated conditions.

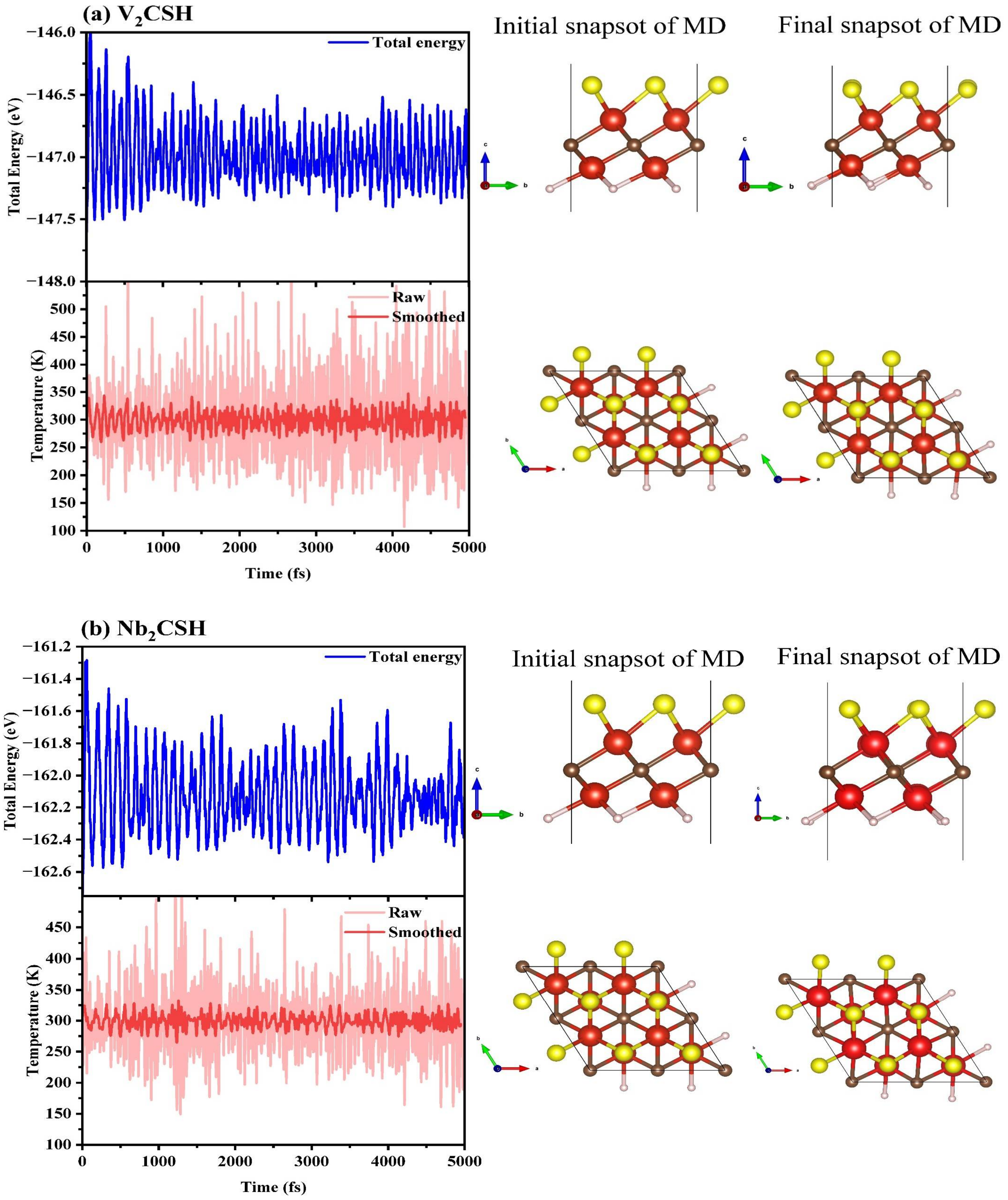

**Figure 5**: Thermal response of Janus (a) $V_2CSH$ and (b) $Nb_2CSH$ monolayers during ab initio molecular dynamics simulations (AIMD) at 300 K. Total energy and temperature trajectories, together with initial and final atomic configurations, show that both monolayer frameworks remain intact throughout the simulations.

## 3.5 Electronic Properties

**Figure 6** presents the element-projected electronic band structures, total and orbital-projected densities of states (DOS), and Fermi surfaces of (a) $V_2CSH$ and (b) $Nb_2CSH$. Both monolayers exhibit metallic electronic structures, as demonstrated by dispersive bands crossing the Fermi level ($E_F$) and a finite DOS at $E_F$. Along the plotted Γ-M-K-Γ path, Fermi-level crossings occur on the Γ-M and K-Γ segments, whereas the M-K segment contains no visible crossing. The participation of multiple bands at $E_F$ establishes the multiband character of the metallic state in both compounds.

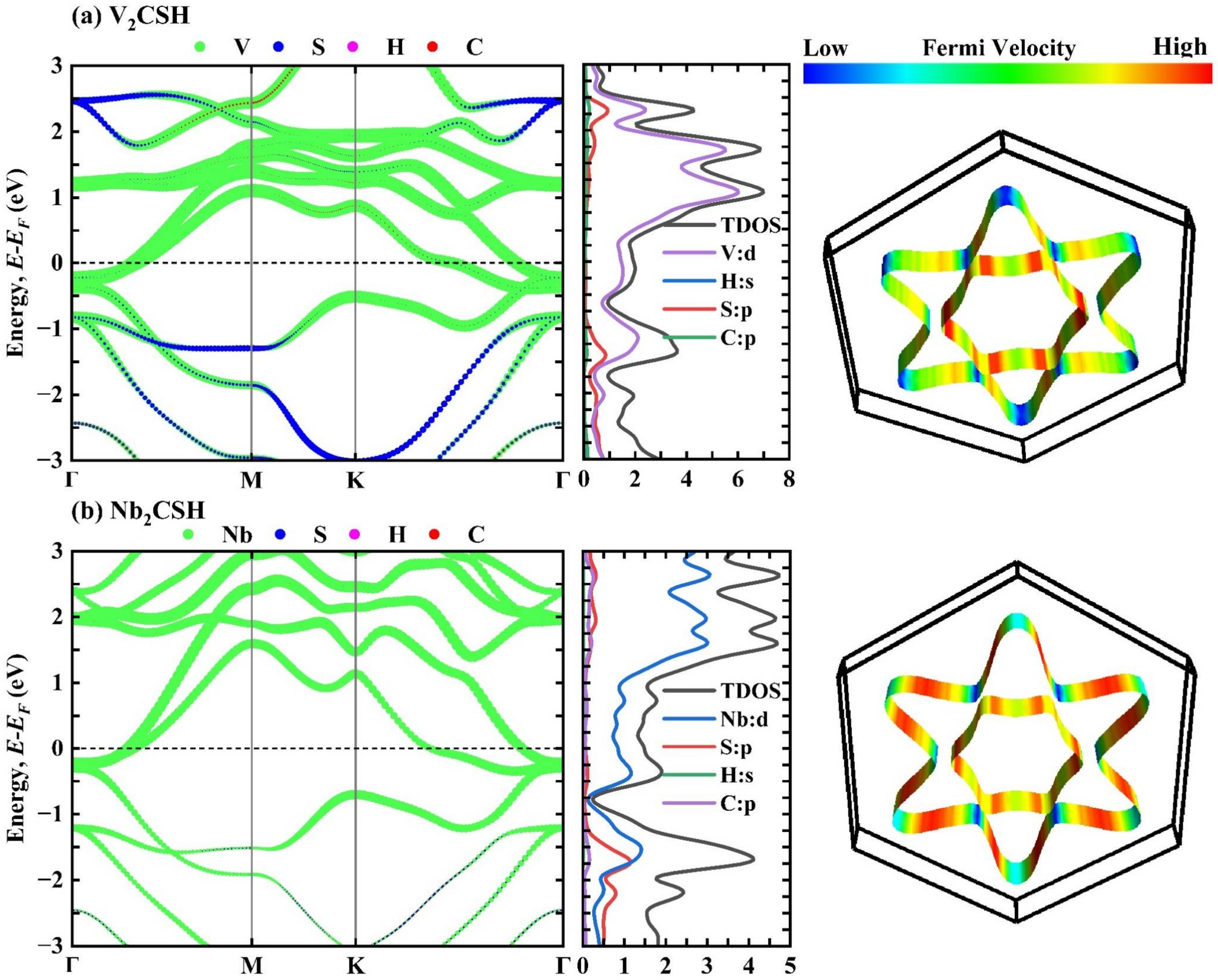

**Figure 6:** Electronic structures and Fermi surfaces of (a) $V_2CSH$ and (b) $Nb_2CSH$ monolayers. Left: Element-projected bands along Γ-M-K-Γ. Middle: Total and orbital-projected DOS, showing dominant V-*d* and Nb-*d* contributions near the Fermi level. Band crossings at $E_F$ = 0 eV (dashed lines) confirm the metallic character of both monolayers. Right: Fermi surfaces colored by Fermi velocity, increasing from blue to red.

The elemental projections identify V and Nb as the main contributors to the bands intersecting $E_F$. The corresponding orbital-projected DOS resolves this contribution into predominantly V-*d* states in $V_2CSH$ and Nb-*d* states in $Nb_2CSH$. Thus, the electronic states available for low-energy excitations arise mainly from the transition-metal *d* manifolds. S-*p* contributions become more apparent in the occupied states below $E_F$, while the C-*p* and H-*s* weights remain comparatively small in its immediate vicinity. The simultaneous V/Nb and S contributions to portions of the occupied bands are consistent with *d*-*p* mixing, although the Fermi-level states retain predominantly *d* character.

The two compounds differ in the energy distribution and dispersion of these *d*-derived states. Within the displayed energy window, several unoccupied branches of $V_2CSH$ lie close together around 1-2 eV above $E_F$, while prominent groups of Nb-derived bands occur at higher energies, around 2-3 eV. Changes are also evident in the occupied-band curvature and the positions of the DOS maxima. Replacing V with Nb therefore modifies the detailed electronic spectrum while retaining metallicity and the dominant *d*-orbital contribution near $E_F$.

Comparable transition-metal *d* dominance has been reported for the related Janus $Ti_2CSH$ monolayer [**26**] and for hydrogen and fluorine-functionalized $V_2C$ MXenes [**46**]. Studies of Ti-based carbide and nitride MXenes further show that surface functionalization can modify *d*-band filling and the resulting Fermi-surface geometry [**47**]. In the present comparison, the common S/H termination is retained, allowing the changes associated with V/Nb substitution to be examined within the same compositional family.

The Fermi-surface renderings in the right panels provide a complementary view of the states crossing $E_F$ throughout reciprocal space. Both compounds display pronounced lobed contours in the central region of the Brillouin zone, producing a star-like appearance in the displayed projection. Their non-circular geometry reflects angular variations of the Fermi wavevectors and the underlying in-plane band dispersion. $V_2CSH$ and $Nb_2CSH$ retain a similar overall contour geometry, with differences in local curvature and in the distribution of colors over the surfaces.

The color scale represents the magnitude of the Fermi velocity, with blue and red indicating lower and higher velocities, respectively. The variation in color along the Fermi surfaces therefore reveals a momentum-dependent velocity magnitude: blue regions correspond to smaller local energy gradients, whereas yellow-to-red regions correspond to larger gradients. This variation distinguishes the electronic dynamics along different portions of the Fermi contours even when their overall shapes are similar. The predominantly *d*-derived states identified at $E_F$ also provide the electronic states participating in the electron-phonon coupling processes examined in the subsequent coupling analysis.

## 3.6 Phonon Spectra and Vibrational Signatures

The phonon dispersions of (a) $V_2CSH$ and (b) $Nb_2CSH$ in **Figure 8** show no imaginary branches along the sampled reciprocal-space paths, supporting the harmonic dynamical stability of both monolayers. Each primitive cell contains five atoms and therefore supports 15 phonon branches, comprising three acoustic and 12 optical branches. Near the Γ point, the longitudinal acoustic (LA) and transverse acoustic (TA) modes predominantly involve in-plane motion, whereas the flexural acoustic (ZA) mode has an out-of-plane character.

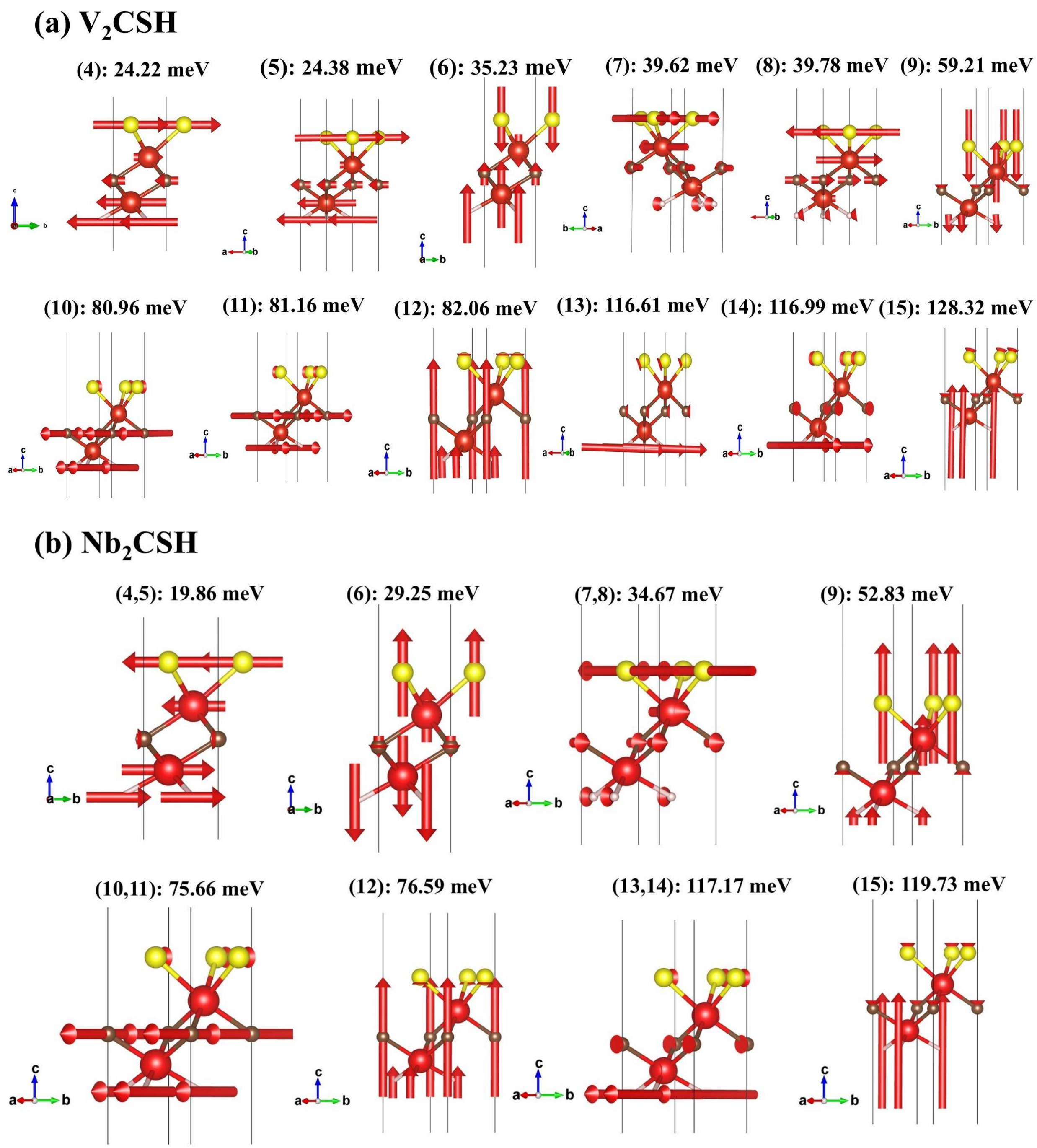

**Figure 7**: The visualization of the Γ-point optical phonon modes of (a) $V_2CSH$ and (b) $Nb_2CSH$, corresponding to the modes reported in **Table 2** and **Table 3**. The displacement patterns illustrate the relative motions of the constituent atoms within the unit cell, elucidating the vibrational character and symmetry of each phonon mode.

At the Γ point, the vibrational modes are classified according to the $C_s$ (m) point-group symmetry of $V_2CSH$ and the $C_{3V}$ (3m) symmetry of $Nb_2CSH$. The assignments in **Table 2** and **Table 3** yield optical representations of 8A′ + 4A″ and 4$A_1$ + 4E, respectively. Thus, $V_2CSH$ has 12 nondegenerate optical modes, whereas $Nb_2CSH$ exhibits four nondegenerate $A_1$ modes and four doubly degenerate E pairs, giving eight distinct optical frequencies. The E pairs describe in-plane vibrations, while the $A_1$ modes correspond to out-of-plane motion in the $Nb_2CSH$ eigenvector visualizations. For $V_2CSH$, A′ and A″ denote even and odd parity under the mirror operation. The separate A′ and A″ frequencies are consistent with the absence of symmetry-protected E-type degeneracy in $C_s$.

For $V_2CSH$, modes 4 (A″) and 5 (A′), at 24.22 and 24.38 meV, respectively, involve collective in-plane displacements of the metal-carbon framework and the surface terminations. Mode 6 (A′) at 35.23 meV is predominantly out-of-plane, involving the metal and termination layers. Modes 7 (A″) and 8 (A′), at 39.62 and 39.78 meV, are characterized mainly by in-plane S vibrations, accompanied by framework displacements. Mode 9 (A′) at 59.21 meV primarily involves out-of-plane motion of S and V atoms. The intermediate-energy modes 10 (A′) and 11 (A″), at 80.96 and 81.16 meV, are predominantly associated with in-plane C vibrations, while mode 12 (A′) at 82.06 meV has a predominantly out-of-plane C character. The highest-energy sector is dominated by H motion: modes 13 (A′) and 14 (A″), at 116.61 and 116.99 meV, involve in-plane displacements, whereas mode 15 (A′) at 128.32 meV corresponds predominantly to out-of-plane H vibrations.

For $Nb_2CSH$, the doubly degenerate E modes 4 and 5 at 19.86 meV describe collective in-plane motion of the framework and termination layers. The $A_1$ mode 6 at 29.25 meV involves out-of-plane displacements, while the E pair 7 and 8 at 34.67 meV is dominated by in-plane S vibrations. The $A_1$ mode 9 at 52.83 meV primarily involves out-of-plane motion of S and Nb atoms. Carbon-dominated vibrations appear in the E pair 10 and 11 at 75.66 meV and the $A_1$ mode 12 at 76.59 meV, corresponding to in-plane and out-of-plane motion, respectively. The E pair 13 and 14 at 117.17 meV is dominated by in-plane H vibrations, and the highest $A_1$ mode, band 15 at 119.73 meV, predominantly involves out-of-plane H motion. These assignments connect the symmetry labels in **Table 3** with the corresponding atomic displacement patterns.

The atom and direction-resolved phonon density of states in **Figure 8** separates the spectra into three principal energy sectors. Metal and S vibrations dominate the low-energy region, extending to approximately 60 meV in $V_2CSH$ and 53 meV in $Nb_2CSH$. Carbon vibrations are concentrated at intermediate energies, approximately 75-90 meV for $V_2CSH$ and 70-82 meV for $Nb_2CSH$, while H vibrations dominate the upper sector, approximately 110-130 meV in both systems. Most corresponding low and intermediate-energy Γ-point modes occur at lower energies in $Nb_2CSH$, consistent with differences in atomic masses and interatomic force constants. The in-plane H modes remain close to 117 meV in both compounds, but the out-of-plane H mode decreases from 128.32 meV in $V_2CSH$ to 119.73 meV in $Nb_2CSH$. This difference indicates that the sensitivity of the termination vibrations to the metal species depends on their direction. The absence of inversion symmetry permits simultaneous Raman and infrared activity. Consistent with the selection rules of the $C_s$ (m) and $C_{3V}$ (3m) point groups, all Γ-point optical modes of $V_2CSH$ and $Nb_2CSH$ listed in **Table 2** and **Table 3** are symmetry-allowed in both Raman scattering and infrared absorption.

**Table 2**: Phonon mode characteristics at the Γ point for the $V_2CSH$, including band indices, symmetry labels, infrared (I) and Raman (R) activity, and corresponding frequencies in meV.

| Bands ν | Subgroup | Active | Energy (meV) |
|---|---|---|---|
| 4 | A″ | I+R | 24.22 |
| 5 | A′ | I+R | 24.38 |
| 6 | A′ | I+R | 35.23 |
| 7 | A″ | I+R | 39.62 |
| 8 | A′ | I+R | 39.78 |
| 9 | A′ | I+R | 59.21 |
| 10 | A′ | I+R | 80.96 |
| 11 | A″ | I+R | 81.16 |
| 12 | A′ | I+R | 82.06 |
| 13 | A′ | I+R | 116.61 |
| 14 | A″ | I+R | 116.99 |
| 15 | A′ | I+R | 128.32 |

**Table 3**: Phonon mode characteristics at the Γ point for the $Nb_2CSH$, including band indices, symmetry labels, infrared (I) and Raman (R) activity, and corresponding frequencies in meV.

| Bands ν | Subgroup | Active | energy (meV) |
|---|---|---|---|
| 4,5 | $E\ L_3$ | I+R | 19.86 |
| 6 | $A_1\ L_1$ | I+R | 29.25 |
| 7,8 | $E\ L_3$ | I+R | 34.67 |
| 9 | $A_1\ L_1$ | I+R | 52.83 |
| 10,11 | $E\ L_3$ | I+R | 75.66 |
| 12 | $A_1\ L_1$ | I+R | 76.59 |
| 13,14 | $E\ L_3$ | I+R | 117.17 |
| 15 | $A_1\ L_1$ | I+R | 119.73 |

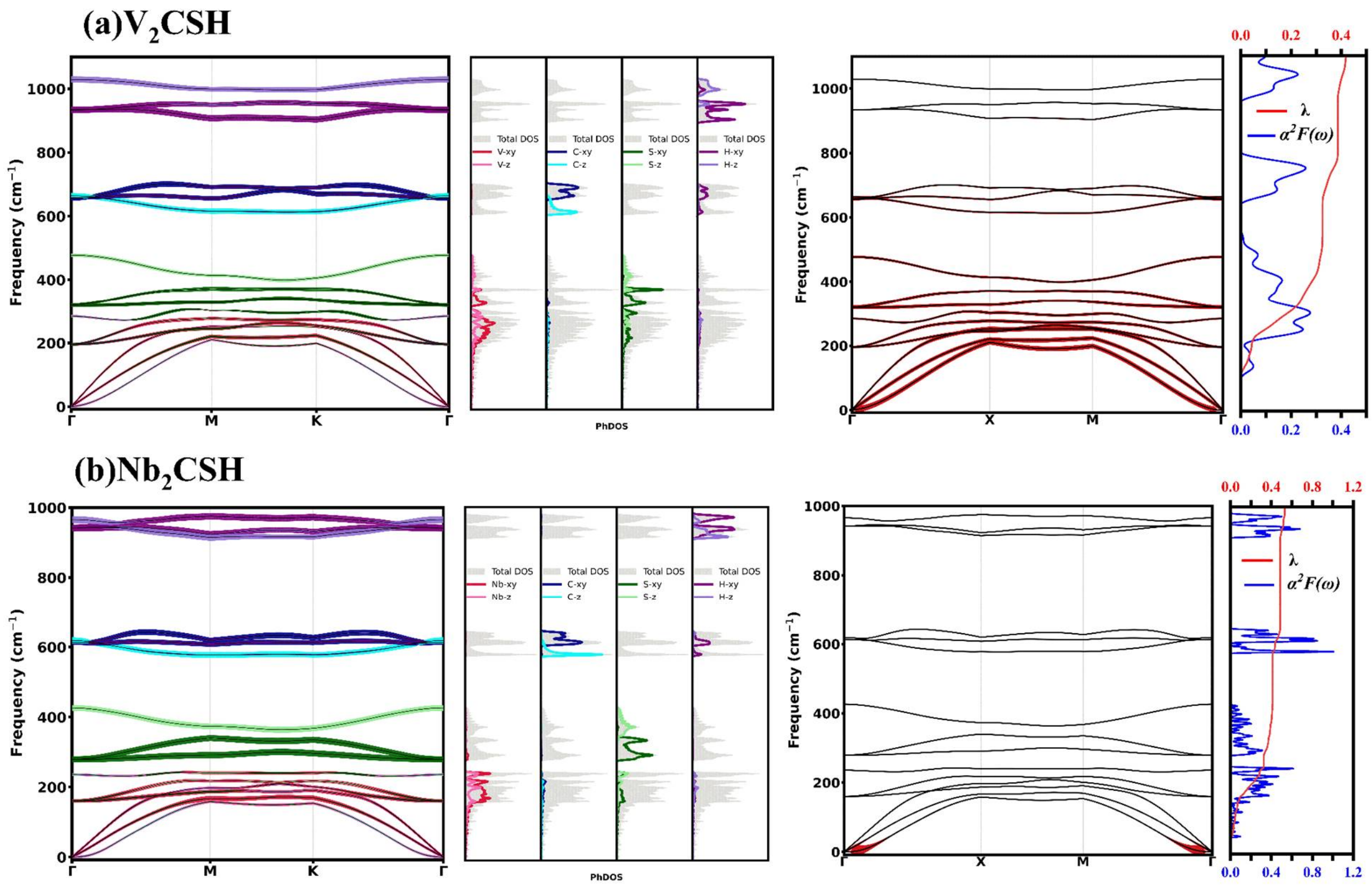

**Figure 8**: Phonon dispersions and electron-phonon coupling (EPC) characteristics of (a) $V_2CSH$ and (b) $Nb_2CSH$. The left panels display phonon dispersions weighted by atomic vibrational contributions, alongside the total phonon density of states (gray shading) and the mode-resolved contributions (colored curves). The right panels present the EPC-weighted phonon dispersions, together with the Eliashberg spectral function, $\alpha^2F(\omega)$, and the cumulative EPC strength, $\lambda(\omega)$.

## 3.7 Electron-Phonon Coupling and Superconductivity

Electron-phonon coupling (EPC) calculations indicate a stronger phonon-mediated superconducting tendency in $Nb_2CSH$ than in $V_2CSH$. The superconducting transition temperature was assessed within the Allen-Dynes modified McMillan formalism [48,49], which relates $T_c$ to the total EPC constant $\lambda$, the logarithmic phonon frequency $\omega_{log}$ and the effective Coulomb pseudopotential $\mu^*$:

$$T_c = \frac{\omega_{log}}{1.2} exp\left[\frac{-1.04(1+\lambda)}{\lambda - \mu^*(1+0.62\lambda)}\right] \tag{10}$$

Here, $\mu^*$ describes the effective screened Coulomb repulsion and is set to 0.10 for both compounds. The Eliashberg spectral function, $\alpha^2F(\omega)$, describes the frequency distribution of EPC and determines the logarithmic average [49]:

$$\omega_{log} = \exp\left[\frac{2}{\lambda}\int \frac{d\omega}{\omega}\alpha^2 F(\omega)\ln\omega\right] \tag{11}$$

The integrated EPC constant is obtained by integrating the spectral function:

$$\lambda(\omega) = 2\int_0^{\infty} \frac{\alpha^2 F(\omega)}{\omega} d\omega \tag{12}$$

The Eliashberg electron-phonon spectral function, $\alpha^2F(\omega)$ is defined as follows [**48**,**49**]:

$$\alpha^2 F(\omega) = \frac{1}{2\pi N(E_F)} \sum_{qv} \frac{\gamma_{qv}}{\hbar\omega_{qv}} \delta(\omega - \omega_{qv}) \tag{13}$$

Here, $\omega_{qv}$ is the phonon frequency, while $\gamma_{qv}$ corresponds to the phonon linewidth for the phonon with wave vector $q$ and branch $v$.

$$\gamma_{qv} = 2\pi\omega_{qv} \sum_{i,j} \int_{BZ} \frac{d^3k}{\omega_{BZ}} \left|g_{qv}(k,i,j)\right|^2 \delta(\varepsilon_{k,i} - \varepsilon_F)\delta(\varepsilon_{k+q,j} - \varepsilon_F) \tag{14}$$

Here, $g_{qv}$ ($k$, $i$, $j$) and $\varepsilon_{k,i}$ represents the matrix of the EPC and the electronic energy, respectively.

The superconducting parameters are summarized in **Table 4**. The coupling constants of 0.39 for $V_2CSH$ and 0.53 for $Nb_2CSH$ place the two compounds in the weak-to-moderate coupling range. The corresponding tabulated transition temperatures are 2.36 and 5.27 K, respectively, identifying $Nb_2CSH$ as the compound with the higher predicted $T_c$.

**Table 4.** Calculated superconducting parameters for $V_2CSH$ and $Nb_2CSH$, including, the electron-phonon coupling constant ($\lambda$), the logarithmic phonon frequency $\omega_{log}$(K) and superconducting transition temperature, $T_c$ (K).

| Compound | $\lambda$ | $\omega_{log}$ | $T_c$ |
|---|---|---|---|
| $V_2CSH$ | 0.39 | 778.35 | 2.36 |
| $Nb_2CSH$ | 0.53 | 357.33 | 5.27 |

Compared with $V_2CSH$, $Nb_2CSH$ exhibits a 35.9% larger $\lambda$, while its $\omega_{log}$ decreases by 54.1%, from 778.35 to 357.33 K. Nevertheless, its tabulated $T_c$ is approximately 2.23 times higher. This trend is consistent with the exponential dependence of $T_c$ on $\lambda$ in the Allen-Dynes expression: in this coupling range, an increase in EPC can compensate for a substantial reduction in the characteristic phonon frequency [**49**]. Thus, the larger $\omega_{log}$ of $V_2CSH$ is insufficient to offset its weaker coupling. The comparison identifies EPC strength as the principal factor favoring the higher transition temperature of $Nb_2CSH$ within the adopted phonon-mediated description.

The distribution of EPC was quantified by partitioning the cumulative coupling into three phonon-energy windows: 0-60 meV, 60-90 meV and above 90 meV. For each compound, the increase in cumulative $\lambda$ within a window was divided by its final cumulative value over the spectral range analyzed. This normalization measures the fractional contribution of each energy sector independently of the total coupling strength. The upper integration limits were approximately 128.21 meV for $V_2CSH$ and 124.07 meV for $Nb_2CSH$.

The low-energy region below 60 meV contributes 76.65% of the integrated EPC in $V_2CSH$ and 77.51% in $Nb_2CSH$. This region is dominated by V and S vibrations in $V_2CSH$ and by Nb and S vibrations in $Nb_2CSH$, as indicated by the atom-resolved phonon density of states in **Figure 8**. The intermediate-energy region, associated predominantly with C vibrations, contributes 14.35% and 14.26%, respectively. The high-energy region above 90 meV, dominated by H vibrations, provides the remaining 8.99% in $V_2CSH$ and 8.23% in $Nb_2CSH$. These percentages describe contributions from frequency windows and their association with particular atoms follows the dominant vibrational character, while individual modes may contain mixed atomic displacements.

The concentration of approximately 77% of the EPC below 60 meV establishes low-energy vibrations as the dominant contribution to $\lambda$ in both compounds. Their importance is consistent with the inverse-frequency weighting of $\alpha^2F(\omega)$ in **Equation 10**, which gives low-frequency spectral weight greater influence on the coupling integral. Carbon and hydrogen-dominated vibrations provide smaller, finite contributions. The comparative analysis reveals that $Nb_2CSH$ achieves a higher $T_c$ due to stronger electron phonon coupling, even though its characteristic phonon frequency is lower. This outcome establishes metal substitution as a decisive factor in tuning superconductivity within Janus MXenes

# 4 Conclusion

This study identifies hydrogenated Janus $V_2CSH$ and $Nb_2CSH$ as stable, nonmagnetic metallic monolayers with phonon-mediated superconductivity. Both compounds retain their asymmetric sulfur/hydrogen termination motif, exhibit isotropic in-plane elasticity and remain structurally robust under finite-temperature simulations. The key distinction lies in their electron-phonon coupling: $Nb_2CSH$ achieves a higher transition temperature (5.27 K) than $V_2CSH$ (2.36 K), despite its lower logarithmic phonon frequency, underscoring coupling strength as the dominant factor.

Approximately 77% of the total coupling originates from low-energy metal-sulfur vibrations, with carbon and hydrogen contributions playing secondary roles. These partitioning highlights the importance of lattice dynamics in governing superconductivity and suggests that targeted manipulation of vibrational modes could further enhance $T_c$. Both the monolayers have relatively small values of $C_{12}$. This implies that a suitable combination of moderate uniaxial pressures may modify the structural and bonding characters significantly, which in turn can have a large impact on superconducting transition temperature and the underlying electronic band structure [**50**-**52**].

Beyond the specific comparison of V and Nb centers, our results establish metal substitution within a fixed Janus termination motif as a general strategy for tuning superconducting behavior and other physical properties in MXenes. Future directions include exploring strain engineering, chemical doping, and external pressure to elevate $T_c$, as well as experimental synthesis routes such as plasma-assisted hydrogenation or chemical vapor deposition. By linking termination asymmetry with superconducting response, this work positions Janus MXenes as a versatile platform for designing next-generation two-dimensional quantum materials.

## Data availability

The data sets generated and/or analyzed in this study are available from the corresponding author on reasonable request.

## Declaration of interest

The authors declare no competing interests.

## CRediT authorship contribution statement

**Jubair Hossan Abir:** Conceptualization, Software, Methodology, Formal analysis, Data curation, Visualization, Writing-original draft, review & editing; **Md Rabi Us Sani Shapon:** Software, Formal analysis, Visualization, Writing-draft, review & editing; **Md. Sharear Aman:** Formal analysis, Visualization, Writing-draft, review & editing; **Md. Kamrul Hassan:** Software, Data curation; **A.K.M. Azharul Islam:** Conceptualization, Supervision, Validation, Administration, Writing- Reviewing and Editing; **Saleh Hasan Naqib:** Conceptualization, Supervision, Validation, Administration, Writing- Reviewing and Editing.

## Appendix A

## Spin-Polarized Electronic Structure without SOC

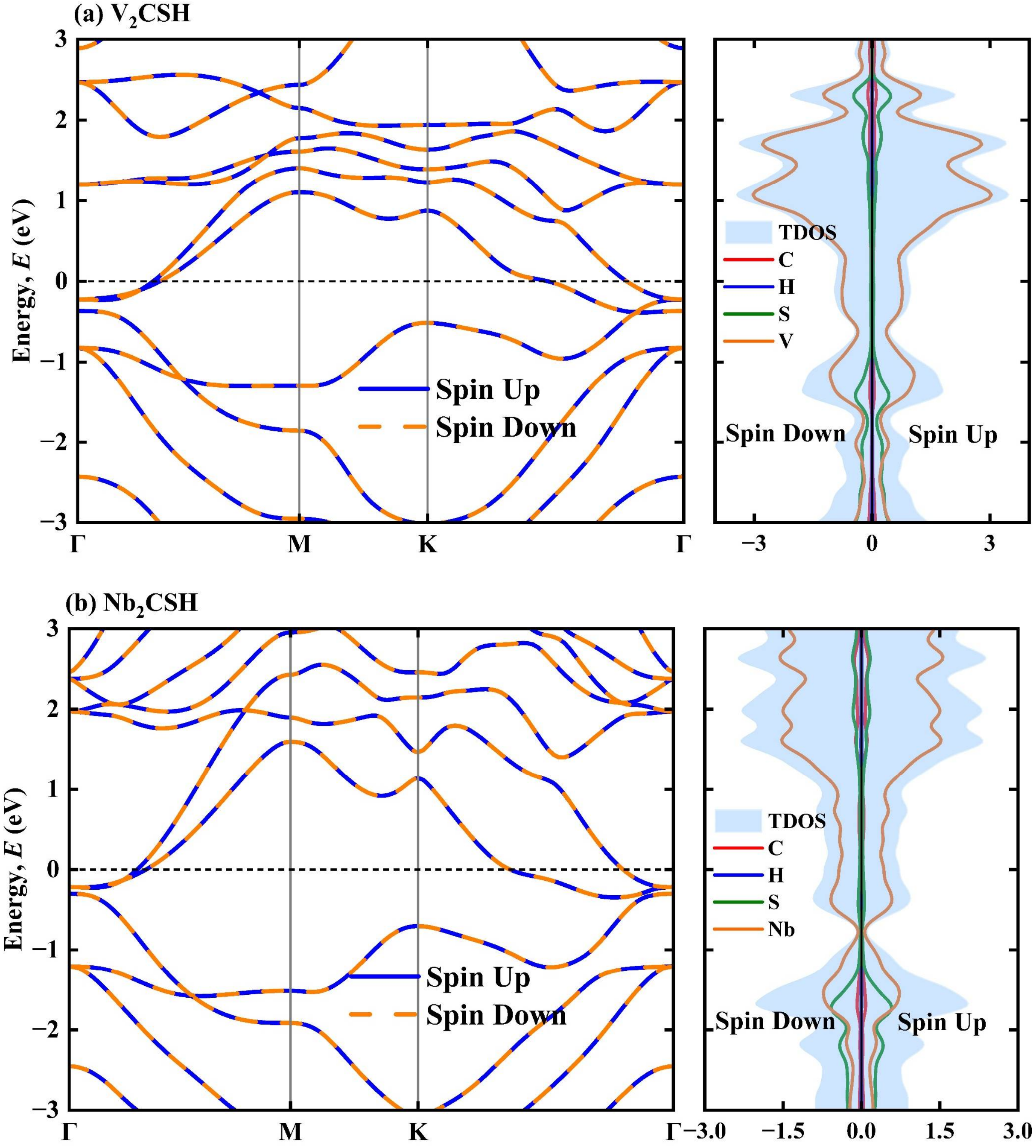

Figure 9: Spin-resolved band structure of monolayer (a) $V_2CSH$ and (b) $Nb_2CSH$. The right-side panel represents Spin-resolved projected density of states (PDOS). The Fermi level was set to 0 eV.